# Variation and Series Approach to the Thomas-Fermi Equation

## M.Oulne

Laboratoire de Physique des Hautes Energies et d'Astrophysique, Faculté des sciences Semlalia B.P.2390, Université Cadi Ayyad, Marrakech 40000, Maroc.

________________________________________________________________________________

**Abstract:**
The Thomas-Fermi equation describing the screening of the Coulomb potential inside heavy neutral atoms is reconsidered. An accurate representation for its numerical solution was obtained by means of the variational principle. The proposed new solution has more precise asymptotic behaviour at large distances from the origin and allows us to obtain the exact value of the initial slope. The obtained new variational solution can also be developed in power series similar to the Baker's ones but more precise even than some series solutions that have been recently obtained within the homotopy analysis method and a modified variational method.
Keywords: Thomas – Fermi equation, Variational principle, series development

________________________________________________________________________________

## I. Introduction.

Due to its simplicity, the statistical model of Thomas-Fermi (TF) knew wide applications in physics [1]. It is considered as a suitable means to estimate the electric charge densities and the electric fields particularly in heavy atoms. This model also allows simplifying many physical problems whose treatment by the method of Hartree - Fock (HF) proves rather complicated. It was the first attempt to relate the energy of a system with its electronic density before the advent of the Density – Functional - Theory (DFT) which was a great success and which is regarded today as being an alternative to the more complicated method of HF. So The TF model has motivated the development of several versions of DFT. It has proved to be efficient in the study of molecules, crystals and Bose-Einstein condensates [2-4]. It has also been applied to atoms in external fields [5] and dense plasmas [6] and astrophysics [7]. But the principal question of this model still remains the solution of its fundamental equation, namely: the TF equation. This famous second order nonlinear differential equation has continued to attract interest of physicists and mathematicians. Several attempts were devoted to solve this problem [8-27] in the past. Recently there has been a renewed interest to solve it both numerically [28] and analytically [29-36]. From the analytical viewpoint various methods have been used to achieve this goal particularly the $\delta$-expansion method [27], the power series development procedure [25], the Pade approximant approach [23], the modified decomposition method [20], the homotopy analysis method [29-32], the variational principle [33-34], the modified variational iteration methods [35] and the Chebyshev pseudospectral approach [36].

Generally, the proposed solutions so far have a bad asymptotic behaviour at long distances from the origin and didn't allow reproducing the numerical value of the initial slope. While the recently obtained solutions have succeeded in reproducing the initial slope with good precision; most of them don't have a precise asymptotic behaviour.

In the present work, we propose a new simple variational solution of the TF equation which reproduces the numerical solution accurately in wide range with a correct asymptotic behaviour at long distances from the origin and which allows us to calculate with exactness the initial slope. The proposed solution will be developed in power series which have the same form as series solutions that have been obtained previously by Baker [23] and recently by Noor [35].

Corresponding author
E-mail address: oulne@ucam.ac.ma

## II. Thomas-Fermi equation.

The Thomas – Fermi non-linear differential equation is [1]:

$$\frac{d^2\phi}{dx^2} = \sqrt{\frac{\phi^3}{x}} \tag{1}$$

where x is a dimensionless variable defined by $x = 4(2Z/9\pi^2)^{-1/3}(r/a_B)$, where r is the distance from the origin, in units of the Bohr radius $a_B$, and Z is the atomic number. The boundary and subsidiary conditions are

$$\phi(0) = 1, \phi(x \to \infty) = 0, \left.\frac{d\phi}{dx}\right)_{x \to \infty} = 0, \tag{2}$$

and

$$\int \rho dv = Z \tag{3}$$

for a neutral atom, where $\rho$ is the electron density which is related to $\phi$ by:

$$\rho = \frac{Z}{4\pi a^3}\left(\frac{\phi}{x}\right)^{3/2} \tag{4}$$

with $a = a_B\left(\frac{9\pi^2}{128Z}\right)^{1/3}$.

The use of the variational principle to the lagrangian [33-34]

$$L(\phi) = \int_0^\infty F dx, \tag{5}$$

where

$$F(\phi, \phi', x) = \frac{1}{2}\left(\frac{d\phi}{dx}\right)^2 + \frac{2}{5}\left(\frac{\phi^{5/2}}{\sqrt{x}}\right), \tag{6}$$

taking into account the constraint (3) is equivalent to the equation (1) [33-34].

## III Results and discussion

In a previous work [34], we proposed the following variational solution

$$\phi(x) = \left(1 + \alpha\sqrt{x} + \beta x e^{-\gamma\sqrt{x}}\right)^2 e^{-2\alpha\sqrt{x}} \tag{7}$$

which depends on three parameters α, β and γ and gave an initial slope equal to - 1.61623647. The error is about 1.77% in comparison with the numerical solution [16]. In order to improve our result, we have to fix the parameter β. The choice of this parameter in the following form

$$\beta = -m\alpha \tag{8}$$

where $0.937 \leq m \leq 1$ allows us to obtain more precise results for the initial slope (table 1).
In table 2, we compare our results for the initial slope with those obtained by Liao [29], Khan [30], Yao [31], Noor [35] and Kobayashi [16]. From this table, one can see that for m = 0.93799968, we reproduce exactly the Kobayashi's result.

In figure 1 we compare the behaviour of our solution given by

$$\phi(x) = \left(1 + \alpha\left(\sqrt{x} - mxe^{-\gamma\sqrt{x}}\right)\right)^2 e^{-2\alpha\sqrt{x}} \qquad (9)$$

where m = 0.93799968 with Khan's solution [30] and the numerical one [24]. Also, we present in the same figure the function [26]

$$\frac{144}{x^3} \qquad (10)$$

known as the Sommerfeld solution which defines the asymptotic behaviour of $\phi(x)$ at large distances from the origin [23]. It is clear that our results are better than those obtained by Khan [30]. The Khan's solution (dashed line) has a bad asymptotic behaviour, while the solution in the present work (dot line) follows correctly the numerical result and asymptotically tends to Sommerfeld solution eq.(10).

The expansion in power series for $\phi(x)$ (equ.(9)) is given by

$$\phi(x) = \sum_{k=0}^{\infty} a_k x^{k/2} \qquad (11)$$

where $a_0 = 1$, $a_1 = 0$, $a_2 = \phi'(0) = B$, $a_3 = 1.358492$ and for $n \geq 4$:

$$a_n = \frac{(-2)^n}{4}\left[\left(\alpha^2 + B\right)\left\{\frac{1}{16}\frac{(\alpha^2+B)(\alpha+\gamma)^{n-4}}{(n-4)!} - \frac{1}{2}\frac{\alpha\left(\alpha+\frac{\gamma}{2}\right)^{n-3}}{(n-3)!} + \frac{\left(\alpha+\frac{\gamma}{2}\right)^{n-2}}{(n-2)!}\right\} + \frac{(4-5n+n^2)\alpha^n}{n!}\right] \qquad (12)$$

The coefficient $a_n$ is a second order polynomial in B.

The obtained eq.(11) is equivalent to Baker's series as cited in Ref.[23] and the series solution recently obtained by Noor [35] but with different coefficients. In table 3, we compare the first 15 power series coefficients in equ.(11) with Baker's [23] and Noor's coefficients [35].

In figure 2 and table 4, we compare the equ.(11) with Baker's series [23], Noor's [35] ones, Khan's results [30] and the numerical solution [24] in the range $0 \leq x \leq 10$.

As seen in table 4 and figure 2, the power series of $\phi(x)$ given in equ.(11) are more precise than Khan's series [30], Baker's ones [23] and Noor's solution [35]. Baker's and Noor's series are out of comparison for $x \geq 1.5$ while Khan's series start to diverge from the numerical solution for x = 4 and don't give good results for $x \leq 0.5$.

Table 1
The parameters of the function (7) and $\phi'(0)$

| m | α | γ | $-\phi'(0)$ |
|---|---|---|---|
| 1 | 0.6057350049 | 0.3715194565 | 1.578384906 |
| 0.938 | 0.6329598887 | 0.3687247046 | 1.58807097 |
| 0.93799968 | 0.6329600376 | 0.3687246935 | 1.588071034 |
| Kobayashi Result [16 ]: -1.588071 | | | |

Table 2
Comparison of $\phi'(0)$ for m = 0.93799968

| Liao[ 29] | Khan[30] | Yao [31] | Noor [35] | Present work |
|---|---|---|---|---|
| -1.58606 | -1.586495 | -1.588005 | -1.588077 | -1.588071 |
| Kobayashi result [ 16 ]: -1.588071 | | | | |

Fig. 1: Comparison between Khan's [30] (dashed line), numerical [24] (solid line) and present work (dot line) results for $\phi(x)$.

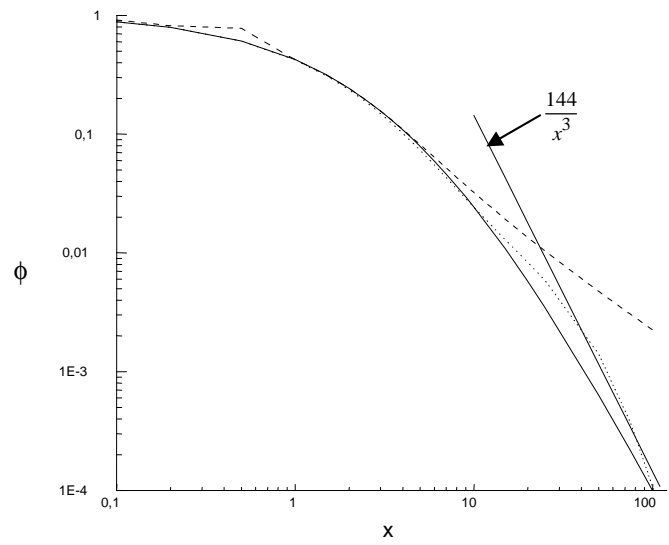

$\dfrac{144}{x^3}$

Table 3
Series coefficients of Baker [23], Noor [35] and the present work

| Coefficients | Baker | Noor | Present work |
| --- | --- | --- | --- |
| $a_0$ | 1 | 1 | 1 |
| $a_1$ | 0 | 0 | 0 |
| $a_2$ | -1.588588 | -1.588077 | -1.588071 |
| $a_3$ | 1.333333 | 1.333333 | 1.358492 |
| $a_4$ | 0 | 0 | -0.005354 |
| $a_5$ | -0.635435 | -0.635231 | -0.873010 |
| $a_6$ | 0.333333 | 0.333333 | 0.915556 |
| $a_7$ | 0.108154 | 0.324256 | -0.585138 |
| $a_8$ | -0.211811 | -0.211744 | 0.279368 |
| $a_9$ | 0.057627 | 0.164041 | -0.107592 |
| $a_{10}$ | 0.014420 | 0.014411 | 0.034840 |
| $a_{11}$ | -0.027132 | 0.006023 | -0.009742 |
| $a_{12}$ | -0.000305 | -0.010172 | 0.002397 |
| $a_{13}$ | 0.146644 | 0.017344 | -0.000527 |
| $a_{14}$ | -0.016784 | -0.016771 | 0.000104 |

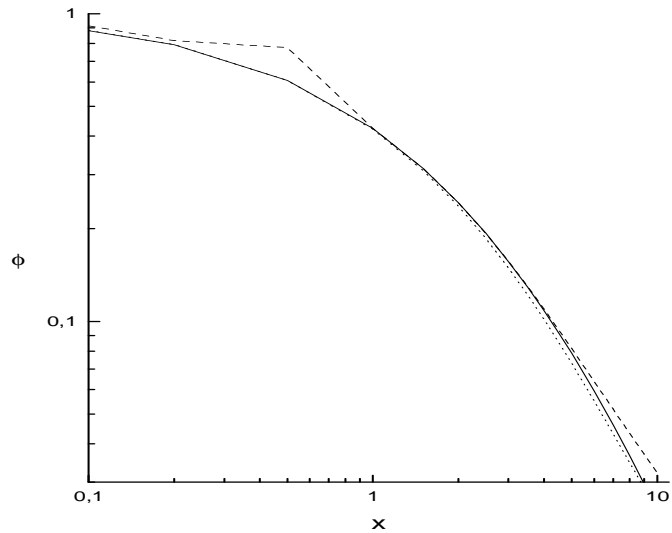

Fig. 2: Comparison of our 30th order series (dot line) with 60[th] order Khan's series (dashed line) and the numerical solution (solid line).

Table 4
Values of $\phi(x)$ for Baker's [23], Noor's [35], Khan's [30] series and the present work (eq.11) in the range $0 \leq x \leq 10$

| X | Num | Baker | Noor | Khan | Present work |
|---|---|---|---|---|---|
| 0.1 | 0.8818 | 0.88164 | 0.88177 | 0.91358 | 0.88209 |
| 0.2 | 0.7931 | 0.79293 | 0.79389 | 0.81809 | 0.79367 |
| 0.5 | 0.6070 | 0.60526 | 0.62992 | 0.77619 | 0.60693 |
| 1 | 0.4240 | 0.41391 | 0.73414 | 0.42377 | 0.42131 |
| 1.5 | 0.3148 | 32.2068 | 1.79283 | 0.31449 | 0.30973 |
| 2 | 0.2430 | 3541.06 | 4.61703 | 0.24272 | 0.23645 |
| 2.5 | 0.1930 | 1.2 E5 | 9.49049 | 0.19279 | 0.18567 |
| 3 | 0.1566 | 2.3 E6 | 14.17561 | 0.15672 | 0.14913 |
| 3.5 | 0.1294 | 2.6 E7 | 10.04827 | 0.12994 | 0.12205 |
| 4 | 0.1084 | 2.2 E7 | -24.4161 | 0.10963 | 0.10152 |
| 4.5 | 0.0919 | 1.4 E9 | -133.642 | 0.09395 | 0.08564 |
| 5 | 0.0788 | 7.2 E9 | -399.313 | 0.08163 | 0.07315 |
| 6 | 0.0594 | 1.3 E11 | - | 0.06382 | 0.05513 |
| 7 | 0.0461 | 1.4 E12 | - | 0.05180 | 0.04309 |
| 8 | 0.0366 | 1.1 E13 | - | 0.04327 | 0.03472 |
| 9 | 0.0296 | 7.1 E13 | - | 0.03700 | 0.02869 |
| 10 | 0.0243 | 3.7 E14 | - | 0.03221 | 0.02421 |

## IV Conclusion

A simple and more precise solution to the Thomas – Fermi equation is obtained by making use of the famous Ritz variational method. Through the comparison among the results of Liao [29], Khan [30], Yao [31], Noor [35] and the present work, it has been shown our work has provided more precise data for the initial slope and the behaviour at large distances from the origin. It has also been shown that the present solution can be developed in power series analogue to the Baker's [23] and Noor's [35] ones but more precise than them. Because of its simple analytical form, the proposed solution in the present work could be used to solve different physical problems based on the Thomas – Fermi model.

## References

[1] Larry Spruch, Pedagogic notes on Thomas-Fermi theory (and on some improvements): atoms, stars and the stability of bulk matter, Rev. Mod. Phys. 63 (5) (1991) 151-209.
[2] D. Ninno, F. Trani, G. Cantele, K. J. Hameeuw, G. Iadonisi, E. Degoli and S. Ossicini, Thomas-Fermi model

of electronic screening in semiconductor nanocrystals, Europhysics. Lett. 74 (3)(2006)519-525.
[3] Y-T Liu, Q-R Zhang, C-Y Gao and W. Greiner ,Thomas–Fermi study on the planar channel fields in crystals, J. Phys. G 29 (3) (2003) L29 – L36 (1).
[4] P. Schuck and X. Viñas, Thomas - Fermi approximation for Bose – Einstein condensates in traps, Phys. Rev. A **61** 43603 (2000).
[5] T. Fennel, G.F. Bertsch and K.-H. Meiwes-Broer, Ionization dynamics of simple metal clusters in Intense fields by the Thomas-Fermi-Vlasov method, Eu. Phys. J. D 29 (3) (2004)367.
[6] V.G. Molinari, M. Sumini and F. Rocchi, Fermion gases in magnetic fields: a semiclassical treatment, Eu. Phys. J. D 12 (2000) 211.
[7] Theodore E.Liolios, The Atomic effects in astrophysical nuclear reactions, Phys. Rev. C 63 (4) (2001)
[8] N. Anderson, A.M. Arthurs and P.D. Robinson, Vriational Solutions of the Thomas – Fermi Equation, Nuovo Cimento 57 (1968)523.
[9] E. Roberts, Upper and Lower Bound Energy Calculations for Atoms and Molecules in the Thomas – Fermi Theory, Phys. Rev. 170 (1968)8-11.
[10] P. Csavinski, Universal Approximate Analytical Solution of the Thomas – Fermi Equation for Ions, Phys. Rev. A 8 (1973)1688-1701.
[11] R.N. Kesarwani and Y.P. Varshni, Improved Variational Solution of the Thomas – Fermi Equation For Atoms, Phys. Rev. A 23 (1981)991-998.
[12] M. Wu, Modified Variational Solution of the Thomas – Fermi Equation for Atoms, Phys. Rev. A 26 (1) (1982)57-61.
[13] B.L. Burrows and P.W. Core, A variational iterative approximate solution of the Thomas – Fermi equation. Quart. Appl. Math. 42 (1984) 73–76.
[14] C.Y. Chan and Y.C. Hon, A constructive method for the Thomas - Fermi equation, Quart. Appl. Math. 44 (1986) 303-307.
[15] A. Cedillo, A perturbative approach to the Thomas - Fermi equation in terms of the density, J. Math. Phys. 34 (1993)2713.
[16] S. Kobayashi, T. Matsukuma, S. Nagai and K. Umeda, Some coefficients of the TFD function, J. Phys. Soc. Japan 10 (1955)759-765.
[17] B.J. Laurenzi, An analytic solution to the Thomas - Fermi equation, J. Math. Phys. 31 (10) (1990) 2535 – 2537.
[18] R.K. Sabirov, Solution of the Thomas - Fermi - Dirac of the statistical model of an atom at small distances from the nucleus, opt. Spect. 75 (1) (1993)1-2.
[19] Y.C. Hon, Adomnian's decomposition method for Thomas - Fermi, SEA Bull. Math. 20 (3) (1996) 55 - 58.
[20] A.M. Wazwaz, The modified decomposition method and the Pade approximants for solving Thomas-Fermi equation, Appl. Math. Comput. J. 105 (1999)11-19.
[21] Tu, Khiet, Analytic solution to the Thomas-Fermi and Thomas-Fermi-Dirac-Weizsäcker equations, J. Math. Phys. 32 (8) (1991)2250-2253.
[22] G.I. Plinkov and S.K. Pogrenya, The analytical solution of the Thomas - Fermi equation for a neutral atom, J. Phys. B 20 (17) (1987)L547.
[23] L. N. Epele, H. Fanchiotti, C.A. Garcia Canal and J.A. Ponciano, Pade approximant approach to the Thomas - Fermi equation, Phys. Rev. A 60 (1) (1999)280-283.
[24] S. Paul Lee and Ta-You Wu, Statistical Potential of Atomic Ions, Chinese J. Phys. 35 (6-11) (1997) 742
[25] M. Francisco M. Fernandez and J.F. Ogilvie, Approximate solutions to the Thomas - Fermi equation, Phys. Rev. A 42 (1) (1991)149-154.
[26] A. Sommerfeld, Asymptotische Integration der Differentialgleichung des Thomas - Fermischen Atoms, Z. Phys. 78 (1932) 283.
[27] C.M. Bender, K.A. Milton, S.S. Pinsky and L.M.JR. Simmons, A new perturbative approach to nonlinear problems, J. Math. Phys. 30 (7) (1989) 1447-1455.
[28] N. A. Zaitsev, I. V. Matyushkin and . V. Shamonov, Numerical Solution of the Thomas – Fermi Equation for the Centrally Symmetric Atom, Russian Microelectronics 33 (7) (2004) 303.
[29] S. Liao, An explicit analytic solution to the Thomas-Fermi equation, App. Math. Comput. 144 (2-3) (2003) 495-506.
[30] H. Khan and H. Xu, Series solution to the Thomas - Fermi equation, Phys. Lett. A 365 (1-2) (2007) 111-115.
[31] Baoheng Yao, A series solution to the Thomas - Fermi equation, App. Math. Comput. 203(1) (2008) 396.
[32] A. El-Nahhas, Analytic Approximations for Thomas – Fermi Equation, Acta Physica Polonica A 114 (4) (2008) 913-918.
[33] Ji-Huan He, Variational approach to the Thomas - Fermi equation, App. Math. Comput. 143 (2003) 533-535.
[34] M. Oulne, Analaytical Solution of the Thomas – Fermi Equation for Atoms, Physics/0511017.

[35] M.A. Noor, S.T. Mohyud-Din and M. Tahir, Modified Variational Iteration Methods for Thomas - Fermi Equation, W. Appl. Sc. J. 4 (4) (2008) 479-486.

[36] K. Parand and M. Shahini, Rational Chebyshev pseudospectral approach for solving Thomas - Fermi equation, Phys. Lett. A 373 (2009)210-213.